\documentstyle[twocolumn,aps]{revtex}
\begin{document} 
\title{
\begin{flushright}
\small
TIFR/CM/99/101(II)
\end{flushright}
Magnetic Phase Diagram of Weakly Pinned Type-II 
Superconductors\\
\vskip 0.25truecm
\normalsize
\noindent
Satyajit Sukumar BANERJEE$^1$, Srinivasan 
RAMAKRISHNAN$^{1,*}$
, Dilip PAL$^1$,
Shampa SARKAR$^1$, Arun Kumar 
GROVER$^{1,*}$,
Gurazada RAVIKUMAR$^2$,
Prasant Kumar MISHRA$^2$, Turumella Venkata 
CHANDRASEKHAR RAO$^2$, Vinod
Chandra SAHNI$^2$, Chakkalakal Varduuny 
TOMY$^3$, Mark Joseph HIGGINS$^4$
and Shobo BHATTACHARYA$^{1,4}$
\vskip 0.1truecm
{\it
$^1$ Department of Condensed Matter Physics and 
Materials Science, Tata Institute of Fundamental 
Research, Mumbai-400005, India\\
$^2$Technical Physics and Prototype Engineering 
Division, 
Bhabha Atomic Research Center, Mumbai-400085, 
India\\
$^3$ Department of Physics, Indian Institute of 
Technology, Powai, Mumbai-400076, India.\\
$^4$ NEC Research Institute, 4 Independence Way, 
Princeton, New Jersey 08540, U.S.A
}
\vskip 0.2 truecm
\small \noindent  
The phenomenon of superconductivity was 
discovered in 1911,
however, the methodology to classify and
distinguish type-II superconductivity was established
only in late fifties after Abrikosov's prediction of a 
flux line
lattice in 1957. The advent of high temperature 
superconductors (HTSC) in 1986 focused attention 
onto identifying
and classifying other possible phases of vortex
matter in all classes of superconductors by a variety
of techniques. We have collated  evidences in support 
of a proposal
to construct a generic phase diagram for weakly 
pinned
superconducting systems, based on their responses to 
ac and dc magnetic fields. The phase diagram 
comprises quasi-glassy phases, like, the Bragg glass, a 
vortex glass and a reentrant glass in addition to the 
(completely) amorphous phases of pinned and 
unpinned variety. The characteristic metastability and 
thermomagnetic history dependent features 
recognized amongst various glassy vortex phases 
suggest
close connections between vortex matter and other
disordered condensed matter systems, like, spin 
glasses,
super cooled liquids/ structural glasses, etc. A novel
quenched random disorder driven fracturing transition
stands out amongst other noteworthy facets  of 
weakly vortex
pinned vortex matter.
\small
\begin{center}
PACS numbers : 64.70 Dv, 74.60 Ge, 74.25 Dw, 
74.60 Ec,74.60 Jg
\end{center}
}
\maketitle
\normalsize
\bigskip
\vskip 0pt
\normalsize
\noindent
\section{Introduction}
Superconducting materials in general are of two 
types, viz., (i) the type-I, which completely shield the 
external magnetic field (H) upto a limiting value called 
the {\it thermodynamic critical field} H$_c$, and (ii) 
the type$-$II, which completely shield the magnetic 
field(H) upto a threshold called the lower critical field 
(H$_{c1}$(T)) and above this field an incomplete 
shielding persists upto an upper critical field 
H$_{c2}$(T), where the type-II superconductor turns 
normal. A phenomenological mean field description of 
the magnetic phase diagram of a type II 
superconductor, as proposed by A. A. Abrikosov 
\cite{rf:1} 
in 1957, comprises a mixed phase between 
H$_{c1}$(T) and H$_{c2}$(T), where the magnetic 
field resides inside the superconductor in the form of 
string like entities called {\it vortices}. Each vortex  is 
associated with a quantum of flux, $\phi_0 = hc/2e $. 
The repulsive interaction between the vortices 
stabilizes them into a regular (generally hexagonal) 
periodic arrangement (often designated as the flux line 
lattice FLL) with an inter-vortex separation $a_0$ 
varying as 
${1/\sqrt H}$, where H is the magnetic field. The 
penetration depth $\lambda$ determines the
range of the repulsive interaction between the 
vortices; the 
interaction falls off as $exp({- r/\lambda}) / \sqrt r$ at 
large distances (r) and it varies logrithmically at small 
distances
\cite{rf:2}. The response of the FLL to stress, is 
similar to that of an elastic medium and one needs to 
associate three elastic moduli for the hexagonal 
symmetry of FLL (e.g., c$_{11}$ (for compression), 
c$_{44}$ (for tilt) and c$_{66}$ (for shear)).

In realistic superconducting samples, the 
superconducting order parameter could get 
preferentially suppressed at locations of atomic 
inhomogeneities, which therefore become 
energetically favorable sites for the localization (i.e., 
pinning) of the normal core of any vortex. A measure 
of the pinning strength experienced by the vortex 
array is obtained via the material attribute {\it critical 
current density ($J_c$)} which can be defined as the 
maximum current that can be sustained by a 
superconductor without depinning the vortices (or 
loosing the zero resistance property). 

Theoretical work of late eighties and nineties has 
shown that by taking into account the effects of 
thermal fluctuations and pinning centers on vortices, 
the mean field description of a type II superconductor 
gets substantially modified and new phases and phase 
boundaries of vortex matter have been predicted 
\cite{rf:3}. In particular, in 1988 D. R. Nelson 
\cite{rf:4} predicted that in a clean pinning free 
system, an ideal FLL phase is stable only in the 
intermediate field range 
under the influence of thermal fluctuations (see 
Fig.1(a) for a schematic plot). A new phase, viz., the 
{\it vortex liquid state}, in which the inter-vortex 
correlation length is of the order of $a_0$, was 
predicted to exist at very low fields near H$_{c1}$ 
and as well as at very high fields just before 
H$_{c2}$, {\it such that the phase boundary marking 
the vortex solid to vortex 
liquid transformation is reentrant. This implies that 
while increasing field at a fixed T, one should first 
encounter the dilute low density($a_0>>\lambda $) 
vortex liquid phase, followed by an ideal FLL and, 
finally, one should reach a very dense
($a_0 \sim \xi $) vortex liquid phase}. Experimental 
studies on high temperature superconductors ({\it 
HTSC}) have established the existence of vortex solid 
to liquid transition at high fields, however, the 
demonstration of the reentrant behavior of melting 
phase boundary  at low fields has so far proved to be 
elusive. The mean field scenario predicts that a 
perfectly periodic arrangement of vortices (ideal FLL) 
should undergo a qualitative transformation under the 
influence of pinning centers such that the vortex 
lattice has spatial correlations only upto a limit as in
a glassy phase \cite{rf:5}. The vortex glass phase can 
exhibit many metastable configurations, each of which 
is characterized by zero linear resistivity. In 1994, T. 
Giamarchi and P. Le Doussal \cite{rf:6} proposed the 
existence of a novel (vortex) solid to solid 
transformation as a function of varying field at a fixed 
temperature. In their framework, a novel Bragg glass 
(i.e., a quasi-FLL) phase at intermediate fields and 
weak disorder transforms (presumably via a first order 
transition) into a vortex glass state at high fields and 
stronger disorder. This solid to solid transformation 
(see Fig.1(b) for a schematic phase diagram) is 
considered to arise due to a proliferation of 
dislocations in the Bragg glass phase, which is initially 
assumed to be free of dislocations of any kind. A large 
fraction  of the experimental efforts since the advent 
of HTSC era have focused on identifying the 
characteristics of different types of dense glassy 
phases of vortex matter\cite{rf:3}. On the other hand, 
relatively little is known about the dilute vortex 
phases under the combined influence of pinning 
centers and thermal fluctuations. A  simulation by 
Gingras and Huse \cite{rf:8} has proposed that the 
addition of pinning can yield a reentrant glass phase at 
low densities (cf. Fig.1(b)), analogous to the low 
density vortex liquid phase (cf. Fig.1(a)) in an ideal 
pinning free situation. 

A fruitful experimental investigation on different 
phases of vortex matter can come about via such 
experimentally accessible quantities, which are related 
to the extent of ordering of the vortices in the various 
phases. A theoretical framework which provides this 
link is the Larkin - Ovchinnikov (LO) description 
\cite{rf:9} propounded in the mid seventies. The LO 
theory showed that the extent of order maintained in 
the FLL in the presence of pinning can be quantified 
in terms of the radial (R$_c$) and the longitudinal 
(L$_c$) correlation lengths, which are distance scales 
over which the deviations {\bf u} of the flux lines 
from their mean periodic lattice positions become of 
the order of the radius of the core of the vortex, i.e., 
the coherence length $\xi$. The R$_c$ and L$_c$, 
which are functions of the pinning strength and elastic 
moduli (which are functions of H and T), are inversely 
related to J$_c$ as, $J_c \propto {1 \over \sqrt 
{R_c^2L_c}}$. It is therefore possible to understand 
the {\it changes in the behavior of J$_c$ in  terms of 
changes in R$_c$ and L$_c$, which could 
happen due to occurrence of phase transformations in 
the vortex matter}.

To enable us to study the evolution of the different 
phases of vortex matter under the combined influence 
of thermal fluctuations and quenched random 
inhomogeneities in the atomic lattice, we have used 
those experimental techniques (namely, ac 
susceptibility, dc magnetization measurements, 
transport experiments, etc.) which provide 
information on the behavior of J$_c$. A remarkable 
feature seen in the behavior of J$_c$ in weakly pinned 
superconducting systems is the phenomenon of Peak 
Effect (PE) \cite{rf:2} prior to reaching the respective 
H$_{c2}$(T) or T$_c$(H). The pinning strength or 
J$_c$ values usually decrease on increasing H or T. 
The PE relates to an anomalous increase in J$_c$, 
terminating in a peak, with increasing H (or T) while 
approaching H$_{c2}$/T$_c$. Despite many years of 
efforts, the PE still awaits complete theoretical 
understanding
\cite{rf:10,rf:11,rf:12}, however it is now widely 
accepted that PE signals the rapid softening of the 
elastic moduli of vortex solid and the occurrence of 
plastic deformation and proliferation of topological 
defects (like, dislocations) in FLL
\cite{rf:12,rf:13}. The vortex array is expected to be 
amorphous at and above the peak position in J$_c$ 
\cite{rf:14}. The phenomenon of PE can therefore be 
exploited to gain information on different phases of 
vortex matter and the processes involved in 
enhancement of loss in order of FLL. We shall 
summarize in this paper the key experimental results 
of PE studies 
\cite{rf:15,rf:16,rf:17,rf:18,rf:19,rf:20,rf:21} in single 
crystals having progressively higher amounts of 
quenched pinning centers and belonging to an 
archetypal low T$_c$ (T$_c$(0) $\sim$ 7 K) weak 
pinning system 2H-NbSe$_2$. These results lead us 
to propose a generic phase diagram for a type II 
superconductor whose details make contacts with 
theoretical predictions summarized schematically in 
Fig.1(b). The proposed phase diagram finds support 
from identical behavior noted in a variety of other low 
T$_c$ superconducting systems 
\cite{rf:16,rf:22,rf:23} and in a single crystal sample 
of a high T$_c$ compound YBa$_2$Cu$_3$O$_7$ 
\cite{rf:24}. A glimpse into some of these results 
\cite{rf:22,rf:23} shall also be provided here.
\section{Experimental}
The ac magnetic susceptibility, dc magnetization and 
dc
resistivity studies have been performed using a 
standard mutual inductance bridge, a Quantum design 
SQUID magnetometer and the usual four probe 
method, respectively. The single crystal samples (A, B 
and C, with progressively increasing number of 
pinning centers) of hexagonal 2H-NbSe$_2$ system 
chosen for illustrating the results here belong to the 
same batches of crystals as utilized by different groups 
of experimentalists in recent years
\cite{rf:25,rf:26,rf:27}. In each of these batches of 
crystals, the FLL with hexagonal arrangement is well 
formed at intermediate fields ($H>0.5 kOe$). The 
single crystals of some other superconducting systems 
such as $CeRu_2$ (T$_c$~=~6.4 K), 
$Ca_3Rh_4Sn_{13}$(T$_c$ = 8.5 K), 
$YNi_2B_2C$(T$_c$ = 15.3 K), etc. studied for 
comparison of trends emerging from data in 2H-
NbSe$_2$ have levels of quenched random 
inhomogeneities comparable to those in sample $C$ 
of 2H-NbSe$_2$.
\section{Results and Discussions}
\subsection{Peak Effect Represents a Phase 
Transition}
We shall first focus attention on an experimental 
result, which amounts to a compelling evidence in 
support of an assertion that PE peak qualifies for the 
status of a phase transition in vortex matter.
Fig.2 depicts the typical variation in the real part of 
the ac susceptibility ($\chi'$) with temperature in 
nominal zero field and in an applied dc field of 4 kOe 
in the cleanest crystal $A$ of 2H-NbSe$_2$ for 
$H_{dc}||c$.
The $\chi '$ curve in zero field provides an estimate of 
the width $\Delta$ T$_c$(0) of the normal-
superconducting transition. We will recall that when 
ac field has fully penetrated the superconducting 
sample, the $\chi '$ response can be approximated in 
Bean's critical state model as \cite{rf:28}:
\begin{equation}
\chi \sim-1+\alpha \frac{h_{ac}}{J_c}; ~~~for~ 
h_{ac} < H^*,
\end{equation}
\begin{equation}
\chi \sim -\beta \frac{J_c}{h_{ac}}; ~~~for~ h_{ac} 
> H^*,
\end{equation}
where $\alpha$ and $\beta$ are geometry and size 
dependent factors and H$^*$ is the field at which 
screening currents flow throughout the sample. Note 
that initially at very low temperatures, $\chi ' \approx -
1$
(perfect shielding), when $h_{ac}$ does not 
penetrate. However, as $h_{ac}$ penetrates the 
sample, the temperature variation in $\chi '$ gets 
governed by the temperature variation in $J_c$. In 
zero field the diamagnetic $\chi '$ response 
monotonically increases (see Fig.2) due to an 
expected decrease in zero field current density 
$J_c(0)$ with the increase in temperature.
However, in a dc field of 4 kOe, which creates a FLL 
with lattice constant 
$a_0 \approx 790 \AA $, the usual decrease in 
diamagnetic $\chi '$(T) response gets interrupted (cf. 
Fig.2) via a sudden onset (at $T=T_{pl}$) of an 
anomalous enhancement in the diamagnetic screening 
response; $\chi '$ reaches a sharp minimum at 
$T=T_p$, above which $\chi '$ recovers rapidly 
towards the normal state value at T$_c$(H). Note 
that the rate of rapid recovery of the latter is very 
much higher than the rate at which the diamagnetic 
$\chi '$ response was decreasing prior to the onset of 
anomalous dip in $\chi '$ at $T=T_{pl}$.
The non-monotonicity in $\chi '$ is a consequence of 
non-monotonicity in $J_c(H, T)$ such that the 
minimum in $\chi '$ corresponds to a peak in $J_c$, 
the ubiquitous peak effect (PE). The experimental fact 
is that the width of the PE region 
is decisively smaller than the width of 
superconducting transition 
($\Delta$ T$_c$(0))
in zero field (cf. Fig.2). It may be further noted (see 
the inset of Fig. 2) that $J_c$ crashes towards a zero 
value from peak position of PE such that the 
diamagnetic 
$\chi '$ response transforms to a paramagnetic 
response across the so called irreversibility 
temperature, $T_{irr}$. Above $T_{irr}$, $J_c 
\approx 0$ and the mixed state of the superconductor 
is in its reversible phase, whose differential magnetic 
response is positive (i.e., diamagnetic dc 
magnetization decreases as dc field/temperature 
increases).
The differential paramagnetic effect (DPE) region in 
the inset of Fig.2, therefore, identifies the depinned 
($J_c$$=0$) state of vortex matter in between 
irreversibility temperature $T_{irr}(H)$ and the 
superconducting temperature T$_c$(H).
To summarize, in a $\chi '(T)$ measurement at a fixed 
H, we can in principle identify four temperatures 
denoted as $T_{pl},~T_p,~T_{irr}~and~T_c$. In the 
subsequent sections, we will present more 
experimental results which elucidate and characterize 
the different phases of vortex matter corresponding to 
$T<T_{pl}$, $T_{pl}<T<T_p$ and 
$T_p<T<T_{irr}$.\\

In weakly pinned systems, $J_c$ is given by the 
pinning force equation:
\begin{equation}
J_cB=(\frac{n_p<f_p^2>}{V_c})^{\frac12}=(\frac{n
_p<f_p^2>}{R_c^2L_c})^{\frac12},
\end{equation}
where n$_p$ is the density of pins, f$_p$ is the 
elementary pinning interaction proportional to the 
condensation energy and V$_c$ is the correlation 
volume of a Larkin domain, within which flux lines 
retain their nominal hexagonal symmetry. As per 
eqn.(3.3), the PE in J$_c$ corresponds to a rapid 
shrinkage in V$_c$ (more rapid than the decrease in 
$<f_p^2>$, which causes the usual monotonic 
decrease in ac screening response with increase in T 
or H), so as to produce an overall enhancement in 
J$_c$ upto the peak position of PE. Above the peak 
position, the rapid collapse in $\chi '$ is expected to 
be governed entirely by a sharp change in elementary 
pinning interaction f$_p$ while approaching the 
superconductor to normal boundary. Transport 
\cite{rf:13} and noise studies \cite{rf:29}
on driven vortex matter in very clean crystals of 2H-
NbSe$_2$ have earlier provided indications that the 
underlying depinned lattice (in the absence of external 
driving force) between the onset and peak positions of 
PE is probably in plastically deformed state in contrast 
to an elastically deformed vortex state existing prior 
to the onset of PE. We shall provide evidences here in 
favor of the scenario that the role of plastic 
deformations (and consequently that of shrinkage in 
V$_c$) ceases at the peak position of PE. It will also 
be elucidated that the onset of shrinkage in V$_c$ is 
induced by the pinning effects in a characteristically 
novel manner.
\subsection{Structure in Peak Effect and 
Identification and Characterization of Phase 
Transformations}
 Fig.3 collates the plots of loci of four features, the 
onset of PE at H$_{pl}$,~T$_{pl}$, the location of 
peak of PE at H${_p}$,~T${_p}$, an apparent 
irreversibility line (H$_{irr}$,T$_{irr}$) and the 
superconducting-normal phase line 
(H$_{c2}$,T$_c$), in crystals A, B and C of 2H-
NbSe$_2$ for fields larger than 1 kOe (i.e., for vortex 
arrays with lattice constants $a_0<1600 \AA$). Note 
that the anomalous PE region spanning from 
(H$_{pl}$, T$_{pl}$) to (H$_{irr}$, T$_{irr}$) 
through (H$_p$, T$_p$) in extremely weakly pinned 
cleanest sample A is so narrow that it suffices for us 
to identify only its H$_p$ line whose thickness 
encompasses its (H$_{pl}$, T$_{pl}$) and 
(H$_{irr}$, T$_{irr}$) data points as well. The PE 
region can be seen to become a little broader in 
weakly pinned crystal B such that the H$_{pl}$, 
H$_p$ and H$_{irr}$ lines in it can be distinctly 
drawn in Fig. 3. However, as the pinning increases 
and one moves onto the nominally weakly pinned 
sample C, the PE region can be seen to expand over 
considerably larger region of (H,T) phase space. It 
can be noted in Fig.3 that the H$_{pl}$ line moves 
away from H$_{irr}$ line as H increases. The 
separation between the H$_{pl}$ and H$_p$ lines 
also correspondingly increases even though H$_p$ 
line remains relatively more close to the H$_{irr}$ 
line.
In the vortex phase diagram of sample C, we have 
labeled the region below the H$_{pl}$ line as the {\it 
elastic glass} and that
between the H$_{pl}$ and H$_p$ lines as the {\it 
plastic glass}. The regions above the H$_p$ line have 
been given the names {\it pinned amorphous}  (for 
H$_p$ $<$ H $<$ H$_{irr}$) and {\it unpinned 
amorphous}  ( for $H_{irr}<H<H_{c2}$), 
respectively. 

As a step towards rational explanation of these 
nomenclatures, we focus attention onto $\chi '$(T) 
behavior (see Figs.4(a) and 4(b)) in sample C in a field 
of 5 kOe ($a_0 \approx 700 \AA $). In Fig.4(a), we 
show a comparison of $\chi '$(T) response in vortex 
states obtained at low temperatures ($T<<T_c$) via 
two entirely different thermomagnetic histories, 
namely, zero field cooled (ZFC) and field cooled 
(FC). Note that in the ZFC case, $\chi '$(T) behavior 
displays a well recognizable PE peak. On the other 
hand, the same feature is far less conspicuous in the 
data recorded during the field cooled warm up (FCW) 
mode. Prior to the peak position of PE, $\chi'$ 
response in the FC state is more diamagnetic than that 
in the ZFC state and above the peak temperature 
T$_p$, the difference between the two responses 
dramatically disappears ({\it a la spin glass 
phenomenon}). The observed behavior is usually 
understood in terms of history dependence in the 
values of J$_c$ (cf. eqn.1), such that J$_c$$^{FC}$ 
$>$ J$_c$$^{ZFC}$ for $T < T_p$ 
\cite{rf:16,rf:20}. 
Eqn.2 implies that history dependence in J$_c$ 
reflects the history dependence in correlation volume 
V$_c$, as neither n$_p$ nor f$_p$, both microscopic 
quantities for the same realization of quenched 
random  disorder, can depend on the thermomagnetic 
history of the vortex system. We have argued earlier 
\cite{rf:16} that in a given circumstance, V$_c$ 
attains the minimum value at the peak position of PE. 
The observation that the diamagnetic $\chi ' $ in FC 
state at $T<<T_p$ is not very different from its value 
at  T = $T_p$ implies that the correlation volume 
$V_c$ at $T<<T_p$ is the {\it supercooled} state 
existing at the peak position of PE. Thus, in the ZFC 
state, the correlation volume of the vortex state prior 
to the arrival of the PE region ($T<<T_p$) is much 
larger than that in the FC state. It can, therefore, be 
concluded that the ZFC vortex state is a well ordered 
one and the FC state at low temperatures attempts to 
freeze in the maximally disordered (i.e., the 
amorphous phase) existing at $T \approx T_p$. Note, 
further, that $\chi'$(T) of the  ZFC state shows two 
sharp ($<1mK~in~width$) jumps at $T \approx 
T_{pl}$ and at $T \approx T_p$. These jumps 
amount to sudden shrinkages in the $V_c$ near the 
onset and peak positions of PE.

The sharp changes in $V_c$ are candidates for first 
order like phase transitions in vortex matter under the 
combined influence of thermal fluctuations and 
pinning centers. To fortify this assertion, we elucidate 
here the occurrence of irreversible behavior while 
thermal cyclings across both the temperatures 
$T_{pl}$ and $T_p$. We show in Fig.4(b) the $\chi 
'$(T) data obtained in sample C in a field of 5 kOe 
while cooling down from temperatures $T_I$, 
$T_{II}$ and $T_{III}$ lying in three different 
regions, viz., (i) $T_I<T_{pl}$, (ii) $T_{pl}<T_{II} 
<T_p$ and (iii) $T_p<T_{III}<T_{irr}$.
The $\chi '$(T) data recorded during warm up of ZFC 
and FC states are indicated by a dotted curve and a 
solid curve (data points omitted), respectively
in Fig.4(b). From this figure, it is apparent that while 
cooling down from $T_I$, $\chi '$(T) response 
retraces its path (i.e., the dotted ZFC curve). 
However, while cooling down from $T_{II}$, the 
$\chi '$(T) response (see filled circle data points) 
attempts to retrace its path while approaching 
$T_{pl}$, but as it nears the temperature $T_{pl}$, 
the $\chi '$(T) response suffers a sharp fall to a 
diamagnetic  
value which lies closer to that of the completely 
amorphous FC like state. On cooling down further 
below $T_{pl}$, the $\chi '$ response never recovers 
on its own  towards the $\chi '$(T) response of  the 
ordered ZFC like state. This striking result can be 
rationalized in terms of the following scenario. When 
an ordered ZFC like state is warmed up towards the 
PE region, the FLL softens, the energy needed to 
create dislocations decreases and the lattice 
spontaneously (partially) fractures at $T_{pl}$.
Upon lowering the temperature from within the PE 
region ($T_{pl}<T<T_p$), the lattice stiffens and the 
stresses build up. The partially disordered vortex 
system fails to drive out dislocations in order to heal 
back to an ordered ZFC like state. Instead, it shatters 
further to relieve its stresses and reaches the 
completely amorphous FC like metastable state. This 
yields an open hysteresis curve in T, i.e., one cannot 
recover the original ordered state by cycling in 
temperature alone. This is a novel feature, not seen in 
the usual first order transitions in which only the 
thermal fluctuations are involved. There are 
compelling evidences that this feature is controlled by 
effective pinning in the system. 
In Fig.4(b), it can finally be noted that while cooling 
down from a temperature $T_{III}$ lying above 
$T_p$, $\chi '$(T) response (see the solid triangle 
data points) faithfully retraces its path while 
approaching $T_p$. Below $T_p$, the cool down 
$\chi '$(T) response starts to differ from that recorded 
earlier during warm up cycles for ZFC and FCW 
states in the same temperature interval. It is clear that 
$d\chi '/dT$ is negative between $T_{pl}$ and $T_p$ 
during warm up cycle (cf. dotted curve in Fig.4(b)), 
whereas it is positive for cool down from $T_{III}$ 
in the same temperature interval. Note, also, that 
eventually 
at $T<T_{pl}$, i.e., after a jump in $\chi '$(T) 
happens while cool down from $T_{II}$, the $\chi 
'$(T) response of both the cool down cycles not only 
nearly overlap but they are also not very different 
from the $\chi '$(T) response recorded during FCW 
cycle. The data of Fig.4(b) thus clearly indicate that 
the transformation occurring in the vortex matter near 
the peak temperature $T_p$ is such that the past 
history of the vortex state is immaterial for its 
behavior above $T_p$. It can therefore be surmised 
that the vortex state is disordered in equilibrium at 
$T>T_p$ and the state of disorder existing at $T 
\approx T_p$ can be preserved or supercooled down 
to much lower temperatures (i.e., $T<<T_p$).
In between $T_{pl}$ and $T_p$, the ordered state 
formed via ZFC (at $T<<T_{pl}$)
mode gets shattered in a stepwise manner. 

The onset and peak positions have special significance 
in the sense that cyclings across them produce the 
unusual open hysteresis curves in T. We have 
demonstrated \cite{rf:15} recently that the first sharp 
jump in $\chi '$ near the onset position is a special 
attribute of the disordering process of FLL which is 
induced by pinning centers as they take advantage of 
the incipient softening of the lattice at PE. The 
connection between the history dependence or 
metastability in vortex lattice can be appreciated by 
examining the $\chi '$(T) data recorded in ZFC and 
FCW modes at the same field (fixed inter-vortex 
spacing $a_0$) in samples with differing quenched 
inhomogeneities (like samples A, B and C of 
$NbSe_2$) or at different fields (i.e., by changing the 
inter-vortex spacing a$_0$) in a given sample. It has 
often been  argued \cite{rf:30} that an increase in field 
(i.e., decrease in $a_0$) amounts to an increase in 
effective pinning as the pinning induced wanderings of 
the flux lines from their mean positions do not 
decrease with field in the same proportion as the 
decrease in $a_0$ with field. Thus, the ratio
$\Delta$$a_0$/$a_0$ (where $\Delta$$a_0$ is the 
pinning induced spread in $a_0$), which can be taken 
as a measure of the effective pinning, increases as $H$ 
increases. We have shown earlier \cite{rf:15} how the 
first jump in $\chi '$ and the associated structure and 
metastability effects in $\chi '$ response between 
$T_{pl}$ and $T_p$ are intimately related. The first 
jump in $\chi '$ can be suppressed and the details of 
structure in $\chi '$ between $T_{pl}$ and $T_p$ can 
be compromised under various circumstances 
\cite{rf:31}, however, the feature of negative peak in 
$\chi '$ at $T_p$ is robust. For instance, the vortex 
state prior to the arrival of PE region can be 
reorganized to a state of even better order (than that 
in the nominal ZFC mode) by subjecting the lattice to 
a sizable driving force  via the process of cyclings in 
large enough ac fields. In such circumstance, the 
behavior of the kind (in ZFC mode) shown in Fig. 
4(a) could transform to that shown in Fig.2, i.e., the 
first jump in $\chi '$ near $T_{pl}$
could be made to disappear, and the detailed structure 
in $\chi '$ smoothened out.\\

A qualitative feature that emerges from the high field 
$\chi '$ data in different crystals of 2H-NbSe$_2$ can 
be summarized as follows. A well correlated (pinned) 
FLL state loses order in a stepwise manner based on 
the competition among three energy scales: the elastic 
energy $E_{el}$, the pinning energy $E_{pin}$ and 
thermal energy $E_{th}$. At low T, $E_{el}$ 
dominates and a phase akin to an elastic (Bragg) glass 
occurs. With increasing T, $E_{el}$ decreases faster 
than $E_{pin}$, and above the (H$_{pl}$,T$_{pl}$) 
line, $E_{pin}$ dominates, leading to proliferation of 
topological defects, likely similar to a vortex glass-
entangled solid phase. At even higher temperatures at 
the (H$_p$,T$_p$) line, the  $E_{th}$ overcomes 
$E_{el}$, and the residual disorder is presumably lost 
due to thermal fluctuation effects (or the effects 
arising out of setting in of divergence in values of 
intrinsic parameters, like, $\lambda$ and $\xi$ near 
superconductor-normal phase boundary). The 
disordered state between (H$_p$,T$_p$) and 
(H$_{irr}$,T$_{irr}$) is pinned with $J_c \ne 0$. 
Finally, a dynamical crossover to unpinned region 
($J_c \approx 0$) occurs above the 
(H$_{irr}$,T$_{irr}$) line and the 
(H$_{c2}$,T$_c$) line identifies the boundary where 
the diamagnetic response in dc magnetization 
measurements falls below a measurable limit.\\ 

\subsection{Reentrant Characteristic in Peak Effect 
Curve and the Effect of Pinning on it}

There is a widespread consensus that the PE curve 
(H$_p$,T$_p$) separates the ordered and disordered 
phases 
of vortex matter. Direct structural support for this 
view point is now also growing via diffraction data 
obtained from small angle neutron scattering 
experiments and via microscopic muon spin rotation 
studies on 
a variety of different superconducting systems, such 
as, pure Nb \cite{rf:14}, 
NbSe$_2$ \cite{rf:32}, (K,Ba)BiO$_3$ \cite{rf:33}, 
etc.. Keeping these in view, we recall that Ghosh {\it 
et al} \cite{rf:27} had shown that T$_p$(H) 
curve determined from $\chi '$(T) at low fields ($<$ 2 
kOe) in sample B of 2H-NbSe$_2$ displayed 
a reentrant characteristic, which bore striking 
resemblance to the theoretically proposed re-entrant 
melting phase boundary shown in the schematic phase 
diagram of Fig.1(a). The turnaround (or the 
reentrance) in 
T$_p$(H) curve in sample B of 2H-NbSe$_2$ has 
been reported \cite{rf:34} for H$\parallel$c and 
H$\parallel$ab at H 
$\sim$ 100 Oe, where FLL constant a$_0$ exceeds 
the range of interaction (i.e., the penetration depth 
values) of vortex lines. The turnaround in T$_p$(H) 
is accompanied by rapid broadening and eventual 
disappearance of PE peak both in magnetic as well as 
transport studies \cite{rf:17}, and such a behavior is 
rationalized by stating that as the inter-vortex 
interaction effects weaken, the collectively pinned 
elastic vortex solid makes a crossover to the small 
bundle or individual pinning regime, which is once 
again dominated by pinning effects of quenched 
random inhomogeneities. The 
reentrant characteristic in T$_p$(H) curve therefore, 
also, implies that as the field gets varied at a fixed 
temperature (isothermal scan), one should first 
encounter the pinning dominated disordered dilute 
vortex state (where $a_0 > \lambda$). This should be 
followed by interaction dominated well ordered phase, 
which should eventually once again transform to a 
disordered (amorphous) phase at very high vortex 
densities ($\lambda >> a_0 \leq \xi$).

A careful examination of the phase diagram shown in 
Fig.3 reveals that for a fixed value of reduced 
temperature t, the onset line (H$_{pl}$, T$_{pl}$) of 
PE progressively moves inwards from sample A to 
sample C. This means that the phase space of ordered 
elastic solid shrinks along the upper portion of the PE 
curve as the pinning effects increase. If the same 
notion is to hold for the lower (reentrant) portion of 
the PE boundary determined first by Ghosh {\it et al} 
\cite{rf:27}, the reentrant leg of the  PE curve should 
move upwards as pinning effects increase. This indeed 
is the experimental situation
\cite{rf:17}. Fig.5 collates the plots of t$_p$(H) 
curves in samples A, B and C in the low field region 
(H$< 1~ kOe$). For the purpose of reference, the 
H$_{c2}$(t) curve has also been drawn for sample A 
in Fig.5 and it may be assumed that H$_{c2}$(t) 
curves for other two samples overlap with that of 
sample A. Note first the error bars on the data points 
of samples B and C as the dc field decreases below 
500 Oe. In sample C, the PE broadens so much that 
the PE peak 
cannot be distinctly identified below about 200 Oe 
\cite{rf:17}. In sample A, however, a sharp PE peak 
(as in Fig.2) can 
be distinctly seen in $\chi '$(T) scans in fields as low 
as 30 Oe. The t$_p$(H) curve in sample A has 
a turnaround feature but its lower (reentrant leg) 
portion could not be determined from temperature 
dependent scans of the in-phase ac susceptibility data. 
However, the trend that the reentrant lower leg of 
t$_p$(H) curve should move upwards, as the pinning 
effects progressively enhance from sample A to C, 
stands 
satisfactorily demonstrated.\\

\noindent
\subsection{Evolution in Pinning Behavior and 
Identification of Reentrant Disordered Region for 
Dilute Vortex Arrays}

In order to more convincing elucidate the reentrant 
characteristic in order to disorder transformation (cf. 
schematics drawn in Fig.1(b)) as the field is increased 
in an isothermal scan, we focus attention onto 
Figs.6(a) to 6(d) and Figs.6(e) to 6(h). These data 
\cite{rf:21} on current densities were extracted from 
an analysis of isothermal in-phase and out-of-phase ac 
susceptibility data as per a prescription of Angurel {\it 
et al} \cite{rf:35} and 
the estimates were verified via dc magnetization 
hysteresis measurements.

Figs.6(a) to 6(d) show J$_c$(H) vs H on log-log plots 
in sample B for four reduced temperatures as 
identified by open circle data points lying on t$_p$(H) 
curve included in the inset of Fig.6(c). Similarly, 
Figs.6(e) to 6(h) depict data in sample A for four 
temperature values marked in the inset of Fig.6(g). 
The two 
sets of plots elucidate the generic nature of evolution 
in shapes of J$_c$(H) vs H behavior. We first focus 
on pairs of plots in Figs.6(a) and 6(d) and in Figs.6(e) 
and 6(h). In both the Figs.6(a) and 6(e), we can 
identify three 
regimes, marked I, II and III in each of them. The 
regime III identifies the quintessential peak effect 
phenomenon. The field region II, in  which J$_c$(H) 
decays with H in a power law manner 
(J$_c$$\sim$H$^{-1}$) identifies the collectively 
pinned elastic vortex solid. The low field region I 
identifies the field span in which the vortices are in a 
small bundle or individual pinning regime and 
consequently J$_c$(H) decays more weakly with H as 
compared to the decay rate seen in the power law 
regime II. Note that the power law region II extends 
down to much lower field values ($\sim$10 Oe) in 
sample A as compared to that ($\sim$200 Oe) in 
sample B. In fact 
in sample B, the J$_c$(H) values move towards the 
saturation J$_c$(H=0) limit of regime I in a faster 
manner than the extrapolated dotted line (of power 
law regime) desires in Fig.6(a). This faster approach 
to the saturated amorphous limit (see Figs.7(a) and 
7(b) for a replot of the data in Figs.6(a) to 6(h) in a 
normalized manner) is in fact the low field counterpart 
of the approach to the peak value of J$_c$(H) at the 
peak position of PE at high fields. Now, we turn to 
the behavior shown in Figs.6(d) and 6(h), where 
J$_c$(H) 
decays very slowly and  monotonically with H (upto 
the highest field values) and the PE region cannot be 
distinctly identified. Note that the (reduced) 
temperatures corresponding to Figs.6(d) and 6(h) in 
samples B and A, 
respectively, are located near the turnaround portion 
of their respective t$_p$(H) curves in the insets of 
Figs.6(c) and 6(g). In an isothermal scan at such a 
temperature, the well ordered collectively pinned 
regime is 
not expected to be encountered. The low field small 
bundle pinning regime is expected to crossover to the 
higher field amorphous vortex phase in a contiguous 
manner. The plots in Figs.6(b) and 6(c) and in 
Figs.6(f) and 6(g) elucidate how the collectively 
pinned well ordered region sandwiched (see arrow 
marks in each 
figures) between the disordered states shrinks as the 
temperature increases while approaching the limiting 
value 
corresponding to the turnaround point of respective 
t$_p$(H) curves.

We collate all the data shown in Fig.6 in two sets of 
normalized plots for samples A and B in Figs.7(a) and 
7(b), respectively. The evolution in shapes of current 
density vs reduced field (H$/$H$_{c2}$(T)) curves in 
the two sets of plots is remarkably similar to the 
evolution in behavior reported earlier in weakly 
pinned 
single crystals of an archetypal low T$_c$ alloy 
superconductor V$_3$Si \cite{rf:36,rf:37} and an 
archetypal high 
T$_c$ cuprate superconductor 
YBa$_2$Cu$_3$O$_7$ \cite{rf:37,rf:38}. This fact 
attests to the generic nature of the 
observed behavior across different classes of 
superconductors. The electromagnetic response of 
type-II materials and the details of the vortex phase 
diagrams in them appear to bear little relationship 
with the microscopics of the different varieties of 
superconducting systems.

The plots in Figs.7(a) and 7(b) clearly bring out how 
the PE broadens as the effects due to interplay 
between the thermal fluctuations and the pinning 
effects enhance while approaching the turnaround 
feature in a PE curve. It can be noted that when the 
PE is very pronounced, J$_c$(H) rises from its 
smallest value at the end of collectively pinned power 
law regime (i.e., at the onset of PE) to reach its 
overall amorphous limit at the 
peak position (see curves from t=0.973 to 0.990 in 
Fig.7(a) and those from t=0.965 to 0.973 in Fig.7(b)). 
At 
the lower field end of the power law regime as well, 
the J$_c$(H) starts to rise faster on lowering the field 
further (see the curve for t=0.965 in Fig.7(b)) and 
reaches the small bundle pinning limit. From the high 
field 
amorphous limit, J$_c$(H) can only decrease on 
increasing the field towards H$_{c2}$ (i.e., b=1) and, 
thereby, it yields a peak effect phenomenon in 
J$_c$(H).
On the otherhand, at the lowest field once the small 
bundle pinning limit is attained, the J$_c$(H) values 
can flatten out towards the J$_c$(0) value. 
This inevitable happening has prompted us recently 
\cite{rf:21} to propose that the said feature be better 
designated as the {\it Plateau Effect} instead of {\it 
reentrant peak effect}, as described earlier 
\cite{rf:17}. Finally, a 
careful examination of the shapes of the curves at 
t=0.983 and 0.99 in Fig.7(a) and that at t=0.977 in 
Fig.7(b) 
reminds us of the phenomenon of second 
magnetization peak or the so called 
{\it fishtail} anomaly widely reported in the context of 
high T$_c$ cuprates in recent years \cite{rf:37,rf:38}. 
It may be useful to mention here that the {\it fishtail} 
effect relates to a characteristic shape of the 
isothermal dc magnetization hysteresis loop. The 
display of data 
in Figs.7(a) and 7(b), therefore, provides a natural 
understanding of the fishtail type anomalous behavior 
in J$_c$(H) in the context of the presence of the 
plateau effect for dilute vortex arrays at low fields and 
the usual peak 
effect for dense vortex arrays at high fields.

We present in Figs.8(a), 8(b) and 8(c) the vortex 
phase diagrams in samples A ,B and C, respectively at 
the low fields and high temperatures (close to 
T$_c$(0)). These phase diagrams have been 
determined from the identification of regimes I, II and 
III via plots as in Fig.6. The lower ends of collectively 
pinned regime II at different temperatures have been 
marked by filled triangle data points, whereas, the 
upper ends of regime II coincide with the dotted 
curve H$_{pl}$(t), which identifies the onset position 
of PE. The span of 
regime I , named as the reentrant disordered 
(following schematics in Fig.1(b)) in Figs.8(a) to 8(c), 
can be identified by slanted lines. The regime III 
corresponding to the PE phenomenon has been 
termed as 
amorphous (see the regions spanned by dotted lines) 
in both the phase diagrams. The so called reentrant 
disordered and amorphous regions overlap and form a 
continuum near the turnaround point of the H$_{pl}$ 
curve. Above the temperature corresponding to the 
turnaround point, the vortex matter remains in a
disordered state at all fields presumably due to 
juxtaposition of effects due to thermal fluctuations 
and pinning. However, at temperatures sufficiently 
below the turnaround temperature, one would find the 
collectively pinned FLL sandwiched between an 
individually pinned disordered low density phase and 
the plastically 
deformed (partially amorphous) high density phase. It 
is apparent that the (H,T) phase space corresponding 
to the collectively pinned ordered state progressively 
shrinks as one goes from the cleanest sample A to the 
nominally pure sample C, however the reenterant 
characteristic in disorder to order transformation 
phenomenon stands elucidated in all the three 
samples. 
\subsection{Vortex Phase Diagrams in Other Low 
T$_c$ Systems}

In earlier sections, we described results pertaining to 
the construction of vortex phase diagrams in weakly 
pinned samples of 2H-NbSe$_2$, which has emerged 
as the most favored superconducting system for 
vortex state studies for its optimum values 
\cite{rf:13} of Ginzburg number G$_i$ and the ratio 
$J_c/J_0$, where J$_0$ is the depairing current 
density. However, in recent years, many reports on 
the construction of vortex phase diagrams have 
appeared in wide classes of other low T$_c$ 
superconducting systems, such as, heavy fermion 
superconductors, UPd$_2$Al$_3$, UPt$_3$, etc., 
\cite{rf:39} mixed valent rare earth
intermetallics, such as CeRu$_2$ \cite{rf:16}, 
Yb$_3$Rh$_4$Sn$_{13}$ \cite{rf:40,rf:41}, etc., 
ternary stannides, like Ca$_3$Rh$_4$Sn$_{13}$ 
\cite{rf:42}, quarternary borocarbide \cite{rf:43} 
superconductors, such as, YNi$_2$B$_2$C, etc. It 
has often been stated that some of the phase 
boundaries (locus of H and T values) drawn for all 
these systems extend over similar parametric limits in 
the normalized (H,T) phase space. Some reports 
\cite{rf:44} have also drawn attention to the 
similarities in the characteristics (in transport studies) 
of different vortex phases identified in systems, like 
CeRu$_2$, YBa$_2$Cu$_3$O$_7$, and 2H-
NbSe$_2$. In a recent study, some of us have shown 
that the vortex phase diagram in a weakly pinned 
single crystal of CeRu$_2$ is nearly identical to that 
in sample C of 2H-NbSe$_2$
\cite{rf:16}. It is our assertion that the weakly pinned 
samples of most other superconducting systems also 
display detailed behavior in the PE region similar to 
that observed in sample C. In isofield ac susceptibility  
studies, the PE region comprises two first order like 
jumps at the onset and the peak positions of PE. In 
between these two jumps, $\chi '$ often displays a two 
peak like structure
\cite{rf:15,rf:16}. Our ansatz is that the first peak 
reflects the commencement of pinning induced 
stepwise shattering of FLL through sudden shrinkage 
of correlation volume V$_c$ at T$_{pl}$ due to 
proliferation of dislocations. The disappearance of 
history dependence above T$_p$ reflects the absence 
of memory of previous history and the complete 
amorphisation of FLL. We present in Figs.9(a) and 
9(b) the $\chi '$(T) data at fixed fields in single 
crystals of two more superconducting systems, 
namely, Ca$_3$Rh$_4$Sn$_{13}$ \cite{rf:22} and 
YNi$_2$B$_2$C \cite{rf:23} in support of the above 
ansatz. In both these figures, the onset (T$_{pl}$) 
and peak (T$_p$) positions of PE phenomenon have 
been marked. Collection of such data along with the 
measurements of isothermal magnetization hysteresis 
loops lead us to the construction of vortex phase 
diagrams in Ca$_3$Rh$_4$Sn$_{13}$ and 
YNi$_2$B$_2$C shown in Figs.10(a) and 10(b), 
respectively. The similarity between these phase 
diagrams and that shown for sample C in Fig.3 amply 
justify the assertion made above. The PE phenomenon 
cannot be observed in crystals of 
Ca$_3$Rh$_4$Sn$_{13}$ and YNi$_2$B$_2$C in 
fields below few kOe, where FLL constant 
a$_0$($\sim$ 1000 $\AA$) exceeds the respective 
penetration depth values and the pinning effects 
dominate over interaction effects. This accounts for 
the termination of H$_{pl}$ and H$_p$ lines at low 
field ends as shown in the Figs.10 (a) and (b).
\section{Conclusion}
The advent of the phenomenon of high temperature 
superconductivity stimulated search for different 
possible phases
(and transformations among them) of vortex matter 
formed as a consequence of competition amongst 
interaction between the vortices and the disorder 
effects caused by thermal fluctuations
and inevitable presence of quenched random 
inhomogeneities in the underlying atomic lattice. 
Theoretical studies \cite{rf:3,rf:4,rf:5,rf:6,rf:7,rf:8} 
during the last decade have brought out how different 
phases of vortex matter could be distinguished on the 
basis of the characteristics of spatial and temporal 
correlations amongst vortices. The phase boundary 
separating the (quasi) Abrikosov flux line lattice and 
the quasi-vortex liquid phases has received the most 
extensive attention from the experimentalists. In the 
field-temperature regions, where thermal fluctuation 
effects dominate over pinning disorder in the context 
of high T$_c$ cuprate systems, magnetic, thermal and 
structural studies have elucidated that the ordered 
vortex solid to the (disordered) vortex liquid 
transition is first order in character 
\cite{rf:45,rf:46,rf:47,rf:48,rf:49}. A crucial evidence 
in favor of this inference has been an observation of a 
step increase in equilibrium magnetization ($\Delta 
M_{eq}$) via bulk and local magnetization 
measurements \cite{rf:45,rf:46,rf:48}, which 
translates into a change in entropy ($\Delta S$) and 
the latent heat $L$ via
the Clausius-Clapeyron relation; $L=T\Delta S=-
T(\Delta M_{eq})
(\delta H_m/\delta T)$. In weakly pinned samples of 
low T$_c$ superconductors, our assertions about the 
different phases of vortex matter and the 
transformations amongst them have been based on the 
characteristic details of the peak effect and their 
relationships to the order to disorder phenomenon. 
The change in equilibrium magnetization that could be 
related to order-disorder transition across the PE is 
difficult to discern experimentally as the PE itself 
manifests as an anomalous increase in irreversible 
magnetization \cite{rf:18}. However, in a limited 
field-temperature region of a very clean crystal of 2H-
NbSe$_2$, Ravikumar {\it et al}\cite{rf:20} have 
revealed the presence of $\Delta M_{eq}$ across the 
PE. The estimated entropy change associated with 
order-disorder transition across PE is of the same 
order as observed earlier in the context of melting of 
FLL in cuprate compounds \cite{rf:20,rf:48,rf:49}. 
Direct structural studies \cite{rf:14,rf:33} across PE 
in weakly pinned samples of low T$_c$ 
superconductors support the notion of occurrence of 
a 
transformation from an ordered (quasi FLL) vortex 
solid to an amorphous vortex phase. Furthermore, 
muon spin mutation studies at low fields in crystals of 
2H-NbSe$_2$ have revealed the occurrence of a 
change in state of vortex matter across PE which is no 
different from the changes reported in the context of 
the fishtail effect and the melting of FLL in similar 
studies in cuprate superconductors \cite{rf:32,rf:50}.

In view of the above, we are tempted to propose a 
generic phase diagram for a realistic sample of a type-
II superconductor which is weakly pinned. We show a 
schematic view of such a phase diagram in Fig.11, 
which has been drawn on the basis of experimental 
data in sample $B$ of 2H-NbSe$_2$. The generic 
phase diagram  comprises six phases, viz., an 
elastically deformed vortex solid phase akin to a 
Bragg glass ($H<H_{pl}$), a plastically deformed 
vortex glass phase ($H_{pl}<H<H_p$), a pinned 
($J_c\ne 0$) amorphous 
phase ($H_p<H<H_{irr}$) and an unpinned ($J_c 
\approx 0$) amorphous phase 
($H_{irr}<H<H_{c_2}$), a low density reentrant 
glass phase ($H_{c_1}<H<H_{low}$) and a 
Meissner phase ($H<H_{c_1}$). The (dotted) 
boundary separating the reentrant glass and the Bragg 
glass phases represents a crossover phenomenon. The 
vortex matter above the PE boundary ($H_p$) is 
disordered in inequilibrium in the sense that no 
metastability effects are observed above this 
boundary. However, the vortex matter can be 
obtained in a disordered metastable state below the 
$H_p$ line by the process of field cooling, which is 
analogous to the phenomenon of supercooling below 
a first order transition. An application of driving force 
can reorganize the disordered metastable state 
towards an ordered stable state \cite{rf:51}. Recent 
experiments \cite{rf:52} have revealed that between 
the $H_{pl}$ and $H_p$ lines, the stationary state of 
the vortex matter is a partially ordered state. In such a 
region, the vortex matter can be obtained in a {\it 
metastable} well ordered state or a {\it metastable} 
amorphous state by the process of superheating or 
supercooling, respectively. An application of a driving 
force could transform both types of metastable 
configurations to the stable partially ordered state.

A phenomenological understanding of transformations 
between the metastable and stable phases of vortex 
mater has also been achieved recently via the 
proposition of a new model \cite{rf:53} which 
reduces to the well known critical state model due to 
C. P. Bean \cite{rf:54} in the absence of metastability 
effects. We
thus believe that the task of sketching a magnetic 
phase diagram of conventional
type-II superconductors stands accomplished 
\cite{rf:55} fairly satisfactorily both from 
experimental view point as well as phenomenological 
(theoretical) considerations.
Many interesting details, however, need to be 
precisely investigated as well 
as quantitatively accounted for.\\

\section{Acknowledgement}

This review is primarily based on the experiments 
performed at Mumbai, India in association with our 
collaborators internationally, who provided us the 
high quality single crystals of a variety of 
superconducting compounds. In particular, we would 
like to acknowledge our cooperation with the groups 
of Prof. Y. Onuki (Osaka, Japan), Prof. D. Mck Paul 
(Warwick,U.K.) and Dr. H. Takeya (Tsukuba, Japan)
and Dr. P. Gammel (Lucent Technoligies, Murray 
Hill, U.S.A.).\\

$Corresponding authors : 
ramky@tifr.res.in/grover@tifr.res.in$

\begin{figure} 
\caption{(a) Schematic vortex phase diagram for an 
ideal
(pinning free) vortex array under the influence of 
thermal
fluctuations (drawn following Ref.3). Note that the 
melting
line separating the Abrikosov flux line lattice (FLL) 
from the
vortex liquid phase has a reentrant character[4].\\
(b) Schematic vortex phase diagram for a vortex array 
under the
combined influence of pinning centers and thermal 
fluctuations (drawn
following Refs. 6-8). Note that the elastically pinned 
FLL (
named as Bragg glass) is sandwiched between the 
dilute and dense
glassy phases of vortex matter named as Reentrant 
glass[8] and vortex glass
[6], respectively. The dotted B$^*$ line separating 
the Bragg glass
and Vortex glass phases is believed to be pinning 
induced, whereas the
solid lines(s) separating the vortex solid phase(s) and 
the vortex liquid
phase is considered to be
caused by thermal fluctuation effects.}
\label{Fig:1}
\end{figure}
\begin{figure}
\caption{ The temperature variation of the in-phase 
($\chi '$) 
ac susceptibility ($h_{ac}$ $\approx$ 1 Oe (r.m.s.) 
and f = 211 Hz) 
in the cleanest crystal A of hexagonal 2H-NbSe$_2$ 
in 
nominally zero bias field (no vortex array) and in a 
dc field of 4 kOe (vortex array with $a_o$$\approx$ 
790$\AA$). 
In zero bias field, the normalized $\chi '$ response 
abruptly increases from -1 towards  near zero value in 
the 
normal state; $\Delta$$T_c$(0) value provides a 
measure
of the width of the superconducting transition. In the 
dc field
 of 4 kOe, $\chi '$ shows a sharp negative peak due
 to peak effect (PE) phenomenon in $J_c$. Note that 
the sharpness 
of PE exceeds that of the superconducting transition. 
The 
inset focuses attention onto a paramagnetic peak 
due to differential paramagnetic effect (DPE) 
located at the edge of the PE peak. Note  
the recipe to locate the positions of onset 
temperature $T_{pl}$ of PE, the temperature of 
PE peak $T_p$, the irreversibility temperature 
$T_{irr}$ and superconducting transition temperature 
$T_c$.}
\label{Fig:2}
\end{figure}

\begin{figure}
\caption{Magnetic phase diagrams (for $H > 1$ kOe) 
determined for 
samples A, B and C of 2H-NbSe$_2$ having 
progressively 
larger number of pinning centers. Note that in sample 
C, the lines marking the onset of PE ($H_{pl}$), the
 onset of reversibility ($H_{irr}$) and the upper 
critical 
field ($H_{c2}$) can be distinctly identified, whereas 
in
 sample A, the $H_{pl}$, the $H_p$ and the 
$H_{irr}$ lines are
 so close that it suffices to draw just the 
$H_p$ line. For justification of nomenclature 
of different phases shown in this figure, see text.}
\label{Fig:3}
\end{figure}

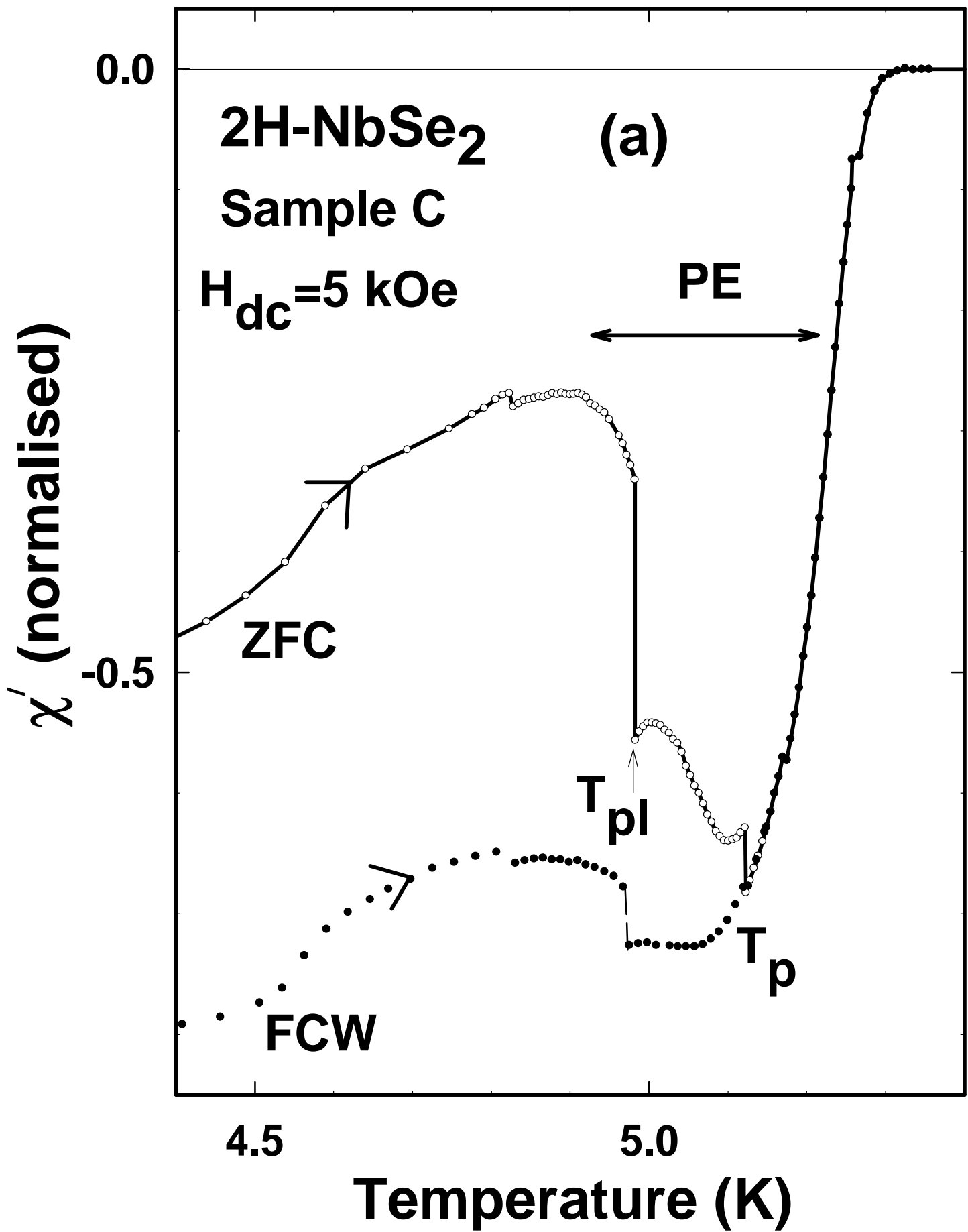
\begin{figure}
\caption{(a) Temperature dependence of $\chi '$ for 
vortex arrays 
created in zero field cooled (ZFC) and field cooled 
(FC) 
modes in a dc field of 5 kOe ($H_{dc}$//c) in sample 
C of
2H-NbSe$_2$. Note the occurrence of two sharp 
changes in
 $\chi '$ response at the onset temperature $T_{pl}$ 
and 
peak temperature $T_p$ in the ZFC mode. Further, 
the 
difference in $\chi '$ behavior between ZFC and 
FC modes disappears above the peak temperature of 
PE
{\it a la spin glass phenomenon}.\\
(b) $\chi '$ responses showing irreversible
 behavior while thermal cycling across the onset 
and the peak positions of PE in sample C in a field of 
5 kOe. The dotted and solid curves show $\chi '$ 
response recorded while warming up the vortex arrays
 created in ZFC and FC modes. The data points refer
 to $\chi '$ behavior recorded while cooling down 
from  (i) $T_I < T_{pl}$ and 
(ii) ($T_{pl}<T_{II}<T_p$ 
and (iii) $T_{III}>T_p$. Note 
that while cooling down from $T_I$, the $\chi '$ 
retraces its behavior recorded while warming up,
 whereas while cooling down from $T_{II}$ and 
$T_{III}$, the $\chi '$ response does not
 retrace its behavior recorded while warming up.}
\label{Fig:4}
\end{figure}
\begin{figure}
\caption{Magnetic phase diagrams (for H $>$ 1 kOe) 
for 
samples A, B and C of 2H-NbSe$_2$. Note that we 
have drawn only the loci of peak temperatures 
in all the three samples. In the field region of 
0.5 to 1 kOe, the PE peak in sample C is also so 
sufficiently narrow that it suffices to draw only 
the locus of peak temperatures in it. However, as 
the field decreases below 500 Oe, the PE peaks in 
sample B and C start to broaden considerably. The 
$t_p(H)$ curve in sample B clearly shows a 
turnaround 
characteristic at H$\sim$ 100 Oe. In sample A, the 
$t_p(H)$ curve can be seen to be on the verge of 
turning around at H$\sim$ 50 Oe.}
\label{Fig:5}
\end{figure}

\begin{figure}
\caption{Log-log plots of $J_c$ vs H at selected 
temperatures in sample B (Figs. 6(a) to 
6(d)) and sample C (Figs.6(e) to 6(h)) of 2H-
NbSe$_2$. 
The peak fields $H_p$ have been marked in Figs.6(a) 
to 6(c)
 and Figs.6(e) to 6(g). The insets 
in Figs.6(c) and in 6(g) show the plots of $t_p(H)$ 
(= $T_p(H)$/$T_c(0)$) and $t_c(H)$ (= 
$T_c(H)$/$T_c(0)$)
 curves in samples B and A, respectively. The data 
points on respective
 $t_p(H)$ curves identify the temperature values at 
which
 the $J_c$ vs H  curves have been displayed in the 
main
 panels. Figs.6(a) and 6(e) show how the entire field 
span 
can be subdivided into three different pinning regimes 
( see text for details). Note that in Figs.6(d) and 6(h), 
the $J_c(H)$ monotonically decays with H and the 
peak
 effect phenomenon cannot be distinctly discerned.}
\label{Fig:6}
\end{figure}

\begin{figure}
\caption{Log-log plots of $J_c(H)$/$J_c(0)$ vs 
H/$H_{c2}$ 
at selected temperatures in sample A (Fig.7(a)) 
and in sample B (Fig.7(b)) of 2H-NbSe$_2$. The 
power law 
regime and the PE region have been marked at the
 lowest reduced temperatures of 0.973 and 0.965 
in samples A and B, respectively. Note that in both
 the samples, the field span over which the power 
law dependence holds reduces as t increases . The 
normalized current  density reaches upto a limiting
 value at the peak of PE. At the lowest field end, 
the normalized current density flattens out to the 
small bundle pinning limit.}
\label{Fig:7}
\end{figure}

\begin{figure}
\caption{Magnetic phase diagram in the low 
field and high temperature region 
in crystals A, B and C of 2H-NbSe$_2$. The H-T 
region between 
the onset of PE (H$_{pl}$ line) and the $H_{c2}$ 
boundary has 
been shaded by dotted lines and the region below 
the onset of power law regime has been shaded by 
solid lines in all the three samples. Note that the  
collectively 
pinned power law regime is sandwiched between the 
so-called
 re-entrant disordered region and the 
amorphous region for 0.97$<$t$<$0.995 in crystal 
A, for 0.95$<$t$<$0.98 in crystal B and for 
0.8$<$t$<$0.95 in crystal C. At temperatures
 above the turnaround features in $H_{pl}$(t) curves,
 the vortex array remains in a disordered state over 
the entire field range.}
\label{Fig:8}
\end{figure}

\begin{figure}
\caption{Temperature variation of $\chi\prime$ for 
vortex
 arrays created in ZFC and FC modes in fields of 
10 kOe and 33 kOe in single crystals of 
$Ca_3Rh_4Sn_{13}$ (Fig.9(a))
 and $YNi_2B_2C$ (Fig.9(b)), respectively. Note the 
occurrence
 of the sharp transition in $\chi\prime$ response in 
ZFC mode 
at temperatures $T_{pl}$ and $T_p$, respectively in 
both 
the crystals. As in Fig. 4(a), the difference in 
$\chi\prime$ behavior between ZFC and FC modes 
disappears above the peak temperature $T_p$.}
\label{Fig:9}
\end{figure}

\begin{figure}
\caption{Magnetic phase diagrams in crystals
 of $Ca_3Rh_4Sn_{13}$ (Fig.10(a)) and 
$YNi_2B_2C$ (Fig.10(b)). 
The nomenclature of different phases shown in these 
diagrams 
follows the prescription justified for 2H-NbSe$_2$ 
system in Fig.3.}
\label{Fig:10}
\end{figure}

\begin{figure}
\caption{A quasi-schematic plot of the magnetic 
phase diagram 
in a weakly pinned type II superconductor (here, for 
example, crystal B of 2H-NbSe$_2$). Different 
vortex states are 
shown sandwiched between $H_{c1}$ and $H_{c2}$ 
lines. This 
diagram has been drawn following analysis of 
experimental 
data on current density in contrast to the schematic 
shown in Fig. 1(b) on the basis of theoretical 
simulations. 
The similarities in the two diagrams attest to the 
appropriateness of the nomenclature of different 
phases in Fig. 11.}
\label{Fig:11}
\end{figure}

\begin{references}
\bibitem{rf:1} A. A. Abrikosov, Sov. Phys. JETP {\bf 
5}  (1957) 1174.
\bibitem{rf:2} Michael Tinkham, Introduction to 
Superconductivity, Mc Graw-Hill, Inc. U.S.A., 
Second edition (1996).
\bibitem{rf:3} G. Blatter, M. V. Feigel'man, V. B. 
Geshkenbein, A. I. Larkin and V. M. Vinokur, Rev. 
Mod. Phys. {\bf 66}  (1994) 1125 and references 
cited therein.
\bibitem{rf:4} D. R. Nelson, Phys. Rev. Lett. {\bf 60}  
(1988) 1973.
\bibitem{rf:5} D. S. Fisher, M. P. A. Fisher and D. A. 
Huse, Phys. Rev. B 
{\bf 43}  (1991) 130. 
\bibitem{rf:6} T. Giamarchi and P. Le Doussal, Phys. 
Rev. Lett. {\bf 72}  (1994) 1530. 
\bibitem{rf:7} T. Giamarchi and P. Le Doussal, Phys. 
Rev. B {\bf 52}  (1995) 1242  and references therein.
\bibitem{rf:8} M. J. P. Gingras and D. A. Huse , 
Phys. Rev. B {\bf 53}  (1996) 15193.
\bibitem{rf:9} A. I. Larkin and Yu. N. Ovchinnikov, 
Sov. Phys. JETP 
{\bf 38}  (1974)854.
\bibitem{rf:10}  A. I. Larkin and Yu. N. Ovchinnikov, 
 J. Low Temp. Phys. {\bf 34}  (1979) 409. 
\bibitem{rf:11} A. I. Larkin, M. C. Marchetti and V. 
M. Vinokur, Phys. 
Rev. Lett. {\bf 75}  (1995) 2992.  
\bibitem{rf:12} C. Tang, X. S. Ling, S. Bhattacharya 
and P. M. Chaikin , 
Europhys. Lett. {\bf 35} (1996) 597.  
\bibitem{rf:13} M. J. Higgins and S. Bhattacharya, 
Physica C {\bf 257}  (1996) 232
 and references cited therein.
\bibitem{rf:14} P. L. Gammel,  U. Yaron,  A. P. 
Ramirez,  D. J. 
Bishop, A. M. Chang, R. Ruel, L. N. Pfeiffer and E. 
Bucher,  
Phys. Rev. Lett. {\bf 80} (1998) 833.
\bibitem{rf:15} S. S. Banerjee, N. G. Patil, S. 
Ramakrishnan, A. K. Grover, S. Bhattacharya, G. 
Ravikumar, P. K. Mishra, T. V. Chandrasekhar Rao, 
V. C. Sahni, M. J. Higgins, C. V. Tomy, G. 
Balakrishnan and D. Mck Paul, Phys. Rev. B, {\bf 59}  
(1999) 6043.
\bibitem{rf:16} S. S. Banerjee, N. G. Patil,  S. Saha, 
S. Ramakrishnan, A. K. Grover, S. Bhattacharya, G. 
Ravikumar, P. K. Mishra, T. V. Chandrasekhar Rao, 
V. C. Sahni, M. J. Higgins, E. Yamamoto, Y. Haga, 
M. Hedo, Y. Inada and Y. Onuki, Phys. Rev. B {\bf 
58}  (1998) 995.
\bibitem{rf:17} S. S. Banerjee, N. G. Patil, S. 
Ramakrishnan, A. K. Grover, S. Bhattacharya, G. 
Ravikumar, P. K. Mishra, T. V. Chandrasekhar Rao, 
V. C. Sahni, M. J. Higgins, C. V. Tomy, G. 
Balakrishnan and D. Mck Paul, Euro. Phys. Lett. {\bf 
44}  (1998) 91.
\bibitem{rf:18} G. Ravikumar, T. V. C. Rao, P. K. 
Mishra, V. C. Sahni, Subir Saha, S. S. Banerjee, N. 
G. Patil, A. K. Grover, S. Ramakrishnan, S. 
Bhattacharya, E. Yamamoto, Y. Haga, M. Hedo, Y. 
Inada and Y. Onuki, Physica C {\bf 276}  (1997) 9.
\bibitem{rf:19} G. Ravikumar, P. K. Mishra, V. C. 
Sahni, S. S.
Banerjee, A. K. Grover, S. Ramakrishnan, P. L. 
Gammel, D. J. Bishop,
E. Bucher, M. J. Higgins and S. Bhattacharya, 
(submitted to Phys. Rev. B); 
http://xxx.lanl.gov/abs/cond-mat/9908222.
\bibitem{rf:20} G. Ravikumar, P. K. Mishra, V. C. 
Sahni, S. S. Banerjee, S. Ramakrishnan, A. K. 
Grover, P. L. Gammel, D. J. Bishop, E. Bucher, M. J. 
Higgins and S. Bhattacharya, Physica C (in press).
\bibitem{rf:21} S. S. Banerjee, S. Ramakrishnan, A. 
K. Grover, G. Ravikumar, P. K. Mishra, V. C. Sahni, 
C. V. Tomy, G. Balakrishnan, D. Mck. Paul, P. L. 
Gammel, D. J. Bishop, E. Bucher, M. J. Higgins and 
S. Bhattacharya, (submitted to Phys. Rev. B); 
http://xxx.lanl.gov/abs/cond-mat/9907111.   
\bibitem{rf:22} Shampa Sarkar, D. Pal, S. S. 
Banerjee, S. Ramakrishnan, A. K. Grover, C. V. 
Tomy, G. Ravikumar, P. K. Mishra, V. C. Sahni, G. 
Balakrishnan, D. Mck. Paul and S. Bhattacharya, 
(submitted to Phys. Rev. B) ; 
http://xxx.lanl.gov/abs/cond-mat/9909297.   
\bibitem{rf:23} D. Pal, Shampa Sarkar, S. S. 
Banerjee, S. Ramakrishnan, A.K.Grover, S. 
Bhattacharya, G. Ravikumar, P. K. Mishra, T.V.C. 
Rao, V. C. Sahni
and H. Takeya, {\it Proceedings of the DAE Solid 
State Physics Symposium}, {\bf 40C} (1997) 310.
\bibitem{rf:24} D. Pal, D. Dasgupta, B. K. Sarma, 
{\it et al}, (unpublished). 
\bibitem{rf:25} S. Bhattacharya and M. J. Higgins, 
Phys. Rev. Lett. {\bf 70}
 (1993) 2617.
\bibitem{rf:26}  W. Henderson, E. Y. Anderi, M. J. 
Higgins and S. Bhattacharya, Phys. Rev. Lett. {\bf 
77}  (1996) 2077.
\bibitem{rf:27} K. Ghosh, S. Ramakrishnan, A. K. 
Grover, Gautam I. Menon, Girish Chandra, T. V. 
Chandrasekhar Rao, G. Ravikumar, P. K. Mishra, V. 
C. Sahni, C. V. Tomy, G. Balakrishnan, D. Mck Paul 
and S. Bhattacharya, Phys. Rev. Lett., {\bf 76}  
(1996) 4600.
\bibitem{rf:28} X. S. Ling and J. Budnick, in 
Magnetic Susceptibility of Superconductors and other 
Spin Systems, edited by R. A. Hein, T. L. Francavilla, 
and D. H. Leibenberg (Plenum Press, New York, 
1991), p. 377. 
\bibitem{rf:29} R. Merithew, M. W. Rabin and M. B. 
Weismann, M. J. Higgins and S. Bhattacharya, Phys. 
Rev. Lett. {\bf 77}  (1996) 3197.
\bibitem{rf:30} G. I. Menon and C. Dasgupta, Phys. 
Rev. Lett. {\bf 73}  (1994)
 1023.   
\bibitem{rf:31} S. S. Banerjee, Ph. D. thesis, 1999. 
University of Mumbai, Mumbai, India.
\bibitem{rf:32} T. V. C. Rao, V. C. Sahni, P. K. 
Mishra, G. Ravikumar, C. V. Tomy, G. Balakrishnan, 
D. Mck Paul, C. A. Scott, S. S. Banerjee, N. G. Patil, 
S. Saha, S. Ramakrishnan, A. K. Grover, S. 
Bhattacharya, Physica C {\bf 299}(1998) 267.  
\bibitem{rf:33} I. Journard, J. Marcus, T. Klein and 
R. Cubitt, Phys. Rev. Lett. {\bf 82}  (1999) 4930.
\bibitem{rf:34} S. S. Banerjee, N. G. Patil, K. Ghosh, 
S. Saha, G. I. Menon, S. Ramakrishnan, A. K. 
Grover, P. K. Mishra, T. V. C. Rao, G. Ravikumar, 
V. C. Sahni, C. V. Tomy, G. Balakrishnan, D. Mck. 
Paul and S. Bhattacharya, Physica B {\bf 237-238}  
(1997) 315.   
\bibitem{rf:35}  L. A. Angurel, F. Amin, M. 
Polichetti, J. Aarts and P. H. Kes, Phys. Rev. B {\bf 
56}  (1997) 3425 and references cited therein.
\bibitem{rf:36} M. Isino, T. Kobayashi, N. Toyota, T. 
Fukase and Y. Muto,
Phys. Rev. B {\bf 38} (1988) 4457.
\bibitem{rf:37} H. K\"upfer, Th. Wolf, C. Lessing, A. 
A. Zhukov, X. LanØon, R. Meier-Hirmer, W. Schauer 
and H. Wuehl, Phys. Rev. B {\bf 58}  (1998) 2886 
and
references cited therein.
\bibitem{rf:38} T. Nishizaki, T. Naito, and N. 
Kobayashi, Phys. Rev. B, {\bf 58} (1998) 11169.  
\bibitem{rf:39} R. Modler, P. Gegenwart, M. Lang, 
M. Deppe, M. Weiden, T. Luhmann,   C. Geibel, F. 
Steglich, C. Paulsen, J. L. Tholence, N. Sato, T. 
Komatsubara,
 Y. Onuki, M. Tachiki and S. Takahashi, Phys. Rev. 
Lett.  {\bf 76}  (1996) 1292.
\bibitem{rf:40} H. Sato,  Y. Akoi,  H. Sugawara and 
T. Fukahara, 
 J. Phys. Soc. Jpn.  {\bf 64} (1995)  3175.
\bibitem{rf:41} C. V. Tomy,  G. Balakrishnan and D. 
Mck. Paul, 
Physica C {\bf 280}  (1997) 1.
\bibitem{rf:42}  C. V. Tomy, G. Balakrishnan and D. 
McK. Paul,
 Phys. Rev. B {\bf 56} (1997)  8346.
\bibitem{rf:43} K. Hirata, H. Takeya, T. Mochiku and 
K. Kadowaki, 
$Advances~in~Superconductivity~VIII$, edited by H. 
Hayakawa and Y. Enomoto (Springer-Verlag, Berlin, 
Germany, 1996) 
p. 619.
\bibitem{rf:44} G. W. Crabtree,  M. B. Maple,  W. K. 
Kwok,  J. Herrmann, 
J. A. Fendrich,  N. R. Dilley and  S. H. Han, Physics 
Essays {\bf 9} (1996)  628
and references cited therein.
\bibitem{rf:45} H. Pastoriza, M. F. Goffman, A. 
Arribre, and F. de la Cruz,
Phys. Rev. Lett., {\bf 72}  (1994) 2951.
\bibitem{rf:46} E. Zeldov, D. Majer, M. 
Konczykowski, V. M. Vinokur
and H. Shtrikman, Nature, {\bf 375}  (1995) 373.
\bibitem{rf:47} A. Schilling, R. A. Fisher, N. E. 
Phillips, U. Welp,
D. Dasgupta, W. K. Kwok, G. W. Crabtree, Nature, 
{\bf 382}  (1996) 791.
\bibitem{rf:48} T. Sasagawa, K. Kishio, Y. Togawa, 
J. Shimoyama,
and K. Kitazawa, Phys. Rev. Lett., {\bf 80} (1998) 
4297 and references therein.
\bibitem{rf:49} M. J. W. Dodgson, V. B. 
Geshkenbein,
H. Nordborg, and G. Blatter, Phys. Rev. Lett., {\bf 
80}  (1998) 837.  
\bibitem{rf:50} T. V. Chandrasekhar Rao, M. J. 
Higgins, V. C. Sahni, G. Ravikumar, P. K. Mishra,   
S. S. Banerjee, N. G. Patil, S. Ramakrishnan, A. K. 
Grover, 
S. Bhattacharya, C. A. Scott, C. V. Tomy, G. 
Balakrishnan and D. McK. Paul (unpublished).
\bibitem{rf:51} S. S. Banerjee, N. G. Patil, S. 
Ramakrishnan, A. K. Grover, S. Bhattacharya, G. 
Ravikumar, P. K. Mishra, T. V. Chandrasekhar Rao, 
V. C. Sahni, M. J. Higgins, Appl. Phys. Lett. {\bf 74}  
(1999) 126.
\bibitem{rf:52} G. Ravikumar, P. K. Mishra and V. C. 
Sahni, S. S. Banerjee, A. K. Grover and S. 
Ramakrishnan, P. L. Gammel, D. J. Bishop and E. 
Bucher,
 M. J. Higgins and S. Bhattacharya, (unpublished).
\bibitem{rf:53} G. Ravikumar, K. V. Bhagwat, V. C. 
Sahni, A. K. Grover, S. Ramakrishnan and S. 
Bhattacharya, (unpublished).
\bibitem{rf:54} C. P. Bean, Phys. Rev. Lett., {\bf 8}  
(1962) 250. 
\bibitem{rf:55} S. S. Banerjee, S. Saha, N. G. Patil, 
S. Ramakrishnan,
A. K. Grover, S. Bhattacharya, G. Ravikumar, P. K. 
Mishra, T. V. C. Rao,
V. C. Sahni, C. V. Tomy, G. Balakrishnan, D. Mck. 
Paul, M. J.Higgins,
Physica C {\bf 308}  (1998) 25.
\end{references}
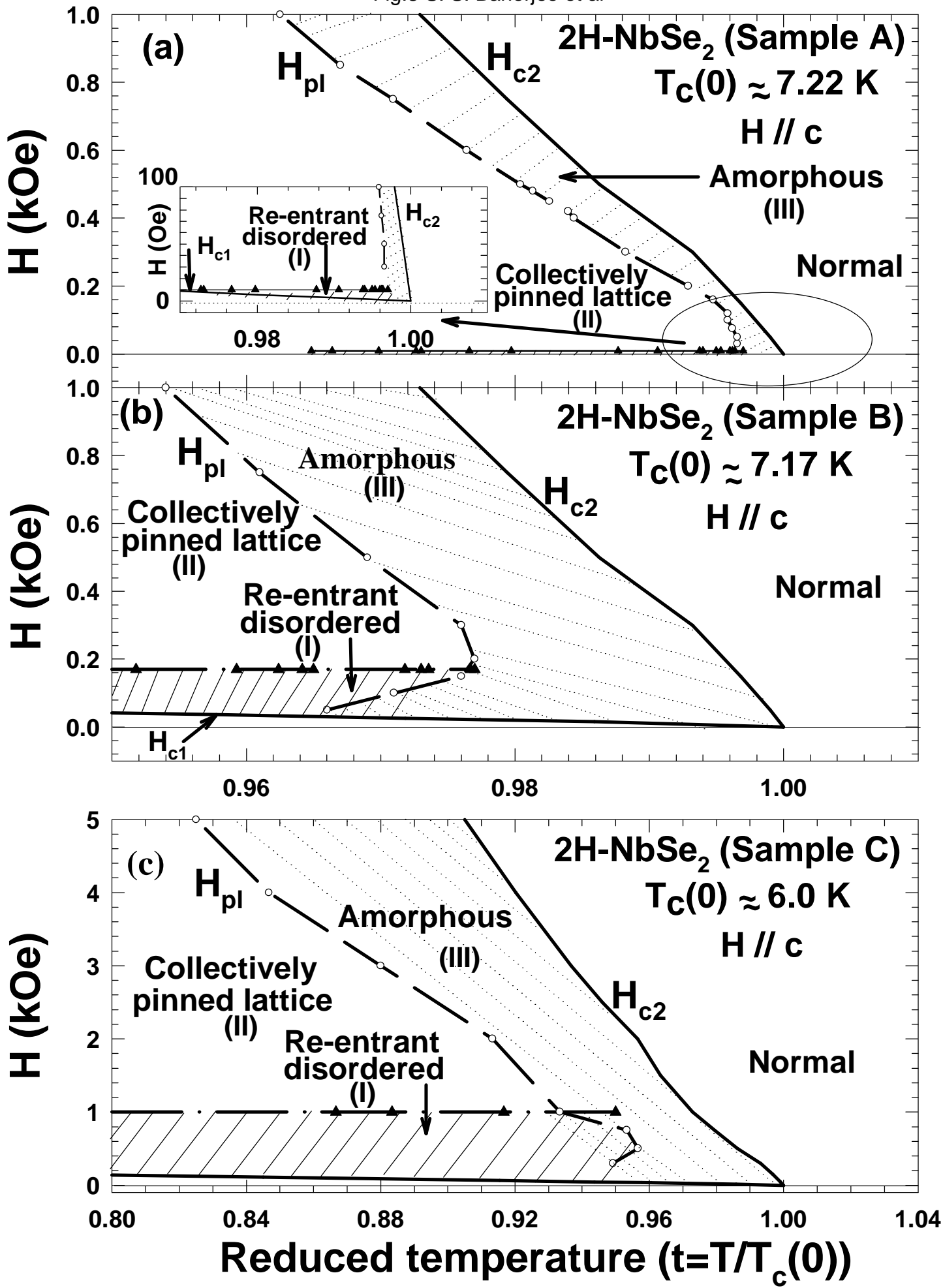
\end{document}